\documentclass[letterpaper, 10 pt, conference]{ieeeconf}

\usepackage{calrsfs,times}
\usepackage{graphicx,psfrag,enumerate}
\usepackage{amssymb}
\usepackage{amsmath}
\usepackage{calrsfs}
\usepackage{color, epsfig, epstopdf, wrapfig}
\usepackage{paralist}
\usepackage{float,cuted}

\newtheorem{theorem}{Theorem}

\newtheorem{lemma}{Lemma} 
 
\newtheorem{assumption}{Assumption}
\newtheorem{problem}{Problem}

\usepackage{color}
\usepackage[normalem]{ulem}
\usepackage{soul} 
\soulregister\cite7
\soulregister\ref7
\soulregister\pageref7

\IEEEoverridecommandlockouts

\begin{document}

\title{\LARGE \bf Detection of Biasing Attacks on Distributed 
  Estimation Networks  
  \thanks{This work was supported by the Australian
  Research Council and the University of New South Wales.}
\thanks{The paper is to appear in Proceedings of the 55th IEEE Conference on
  Decision and Control, Las Vegas, December 2016.}}

\author{Mohammad Deghat \and Valery Ugrinovskii \and Iman Shames \and
  C\'{e}dric Langbort%
\thanks{M. Deghat and V. Ugrinovskii are with the School of Engineering and Information Technology, University of New South Wales at the Australian Defence Force Academy, Canberra, ACT 2600, Australia. {\tt\small m.deghat@unsw.edu.au;v.ougrinovski@adfa.edu.au}}%
\thanks{I. Shames is with the Department of Electrical and Electronic Engineering, University of Melbourne, Melbourne, Victoria 3001, Australia. {\tt\small iman.shames@unimelb.edu.au}}
\thanks{C. Langbort is with the Department of Aerospace Engineering and Coordinated Science Laboratory, University of Illinois at Urbana-Champaign, Urbana, IL 61801 USA. {\tt\small langbort@illinois.edu}}
}

\maketitle
         
\begin{abstract}
The paper addresses the problem of detecting attacks on 
distributed estimator networks that aim to intentionally bias process
estimates produced by the network. It provides a sufficient condition, in
terms of the 
feasibility of certain linear matrix inequalities, which guarantees
distributed input attack detection using an $H_\infty$ approach. 

\end{abstract}

\section{Introduction}
With recent rapid developments in the area of networked control and
estimation
, the security of networked systems against
input attacks and faults becomes increasingly important.
The mainstream of the results in the literature focus on centralized attack
and fault detection, however some recent work has been done on distributed
attack and fault detection due to the fact that not all measurements might
be available at each node of the network; 
see~\cite{Ferrari-2012,PDB-2013,PDB-2015,TSSJ-2014,HWJZ-2011,GHJ-2013} and the
references therein.

This paper considers the problem of detection of attacks on
consensus-based distributed estimation networks. The topic of distributed
estimation has gained considerable attention in the literature, in a bid to
reduce communication bottlenecks and improve reliability and fidelity of
centralized state observers. Filter cooperation and consensus ideas have
proved to be instrumental in the design of distributed state
observers~\cite{Olfati-Saber-2007,SWH-2010,U6}. At the same time,
consensus-based systems are particularly vulnerable to intentional attacks
since the compromised agents can interfere with the functions of the entire
network in a significant way~\cite{PBB-2012}. Uncertainty and noise
represent another challenge from the attack detection viewpoint --- state
observers are typically required in applications where uncertainty and
noise make accessing the system state difficult; this may allow the
attackers to remain undetected by injecting signals compatible with the
noise statistics~\cite{PDB-2013}. This motivates an increased interest in the
literature in detection of rogue behaviours of state observers.
  
In this paper, we consider a general framework of distributed state
estimation considered, for example, in~\cite{SWH-2010,U6,LaU1} and assume
that some of the nodes of the network are compromised. Mathematically, this
situation is modelled by allowing the
compromised observers to be driven by certain attack/fault
inputs. The purpose of the attack under consideration is
to force the compromised node to produce biased state estimates and then
exploit the consensus mechanism within the network to propagate those
estimates across the network. 
Conventional false data injections into measurements can also be included
in the model as a routine extension of our results. 

From the viewpoint of fault detection/input estimation, the system subject
to attack is distributed itself. This is similar to
\cite{TSSJ-2014}, but is different from 
\cite{HWJZ-2011,GHJ-2013} which were focused on detecting faults applied to
the observed plant. 
We use an
$H_\infty$ fault detection approach which allows for a broad range of
uncertainty in the sensors and the plant model, as well as a quite broad
range of attack inputs. Furthermore, to detect
the attack/fault, the proposed attack observers use the same 
plant measurements and the state estimate information communicated from the
neighbours as the state observers themselves. The key idea is to use this
information, without additional communication overheads, to determine which
of the node observers' behaviour differs from what this information predicts.   

Our idea of governing the detectors by neighbours' state estimates to track
the attack input is similar to~\cite{SA-2015}, where integral action
controllers governed by diffusive couplings were used for averaging
constant disturbances. More precisely, in~\cite{SA-2015} distributed
integral action controllers were used for averaging constant disturbances
to enable all agents in the system to synchronize to a common reference
system governed by the averaged constant disturbance. In contrast, here we
are interested in tracking individual attack inputs, rather than tracking
an averaged attack vector. Technically this required us to introduce
additional dynamics into the fault detectors. Also unlike~\cite{SA-2015},
the $H_\infty$ formulation adopted here does not restrict the attack inputs
to be constants. 

The paper is organised as follows. In
Section~\ref{sec:distributed_estimation}, a background on distributed
consensus based estimation is presented. Also, the idea of
distributed attack estimation with $H_\infty$ consensus is explained and
the attack detection problem is formulated in that section. The main result
is given in Section~\ref{sec:lmi_design}, where a sufficient condition in
terms of coupled linear matrix inequalities is expressed. Concluding remarks are given in Section~\ref{sec:conclusion}. 

\emph{Notation}: $\mathbf{R}^n$ denotes the real Euclidean $n$-dimensional vector space, with the norm  $\|x\|=(x'x)^{1/2}$; here the symbol $'$ denotes the transpose of a matrix or a vector.
The symbol $I_n$ denotes the $n\times n$ identity matrix, and $0_{m\times n}$ denotes the zero matrix of size $m\times n$. We will occasionally use $I$ and $0$ for notational convenience if no confusion is expected. For real symmetric $n\times n$ matrices $X$ and $Y$, $Y>X$ (respectively, $Y\geq X$) means the matrix $Y-X$ is positive definite (respectively, positive semidefinite).
The notation $L_2[0, \infty)$ refers to the Lebesgue space of $\mathbf{R}^n$-valued vector-functions $z(.)$, defined on the time interval $[0, \infty)$, with the norm $\|z\|_2\triangleq\left(\int_0^\infty \|z(t)\|^2 dt \right)^{1/2}$ and the inner product $\int_0^\infty z_1'(t) z_2(t) dt$.
   
\section{Formulation of the distributed attack detection problem}
\label{sec:distributed_estimation}
\subsection{Network topology}
Consider a filter  network with $N$ nodes and a directed graph topology $\mathbf{G} = (\mathbf{V},\mathbf{E})$ where
$\mathbf{V}$ and $\mathbf{E}$ are the set of vertices and the set of edges (i.e, the subset of the set $\mathbf{V}\times \mathbf{V}$), respectively. 
Without loss of generality, we let $\mathbf{V}=\{1,2,\ldots,N\}$. 
The graph $\mathbf{G}$ is assumed to be directed, reflecting 
the fact that while node $i$ receives the information from node $j$, this
relation may not be reciprocal
. The notation $(j,i)$ will denote the edge
of the graph originating at node $j$ and ending at node $i$. It is assumed that
the nodes of the graph $\mathbf{G}$ have no self-loops, i.e.,
$(i,i)\not\in \mathbf{E}$. 
   
For each $i\in \mathbf{V}$, let $\mathbf{V}_i=\{j:(j,i)\in \mathbf{E}\}$ be the set of nodes supplying information to node
$i$. 
The cardinality of
$\mathbf{V}_i$, known as the in-degree of node $i$, is denoted $p_i$; i.e.,
$p_i$ is equal to the number of incoming edges for node $i$. Also,  $q_i$
will denote the number of outgoing 
edges for node $i$, known as the out-degree of node $i$.
Let $\mathbf{A}=[\mathbf{a}_{ij}]$ be the adjacency matrix of the
digraph $\mathbf{G}$, i.e., $\mathbf{a}_{ij}=1$ if $(j,i)\in \mathbf{E}$,
otherwise $\mathbf{a}_{ij}=0$.
Then, $p_i=\sum_{j=1}^N\mathbf{a}_{ij}=\sum_{j\in \mathbf{V}_i} \mathbf{a}_{ij}$, $q_i=\sum_{j=1}^N\mathbf{a}_{ji}$. 

\subsection{Background: distributed consensus-based $H_\infty$ 
  estimation} 
A typical distributed consensus-based $H_\infty$ 
  estimation problem considers a plant described  by the equation
\begin{equation}
  \label{eq:plant}
  \dot x=Ax+B_2\xi(t), \quad x(0)=x_0, \quad x\in\mathbf{R}^n,
\end{equation}
governed by an disturbance input $\xi\in \mathbf{R}^m$. A network of
filters connected according to the graph $\mathbf{G}$ takes measurements of
the plant with the purpose to produce an estimate of $x$. It is assumed
that each filter takes measurements   
\begin{equation}\label{U6.yi}
y_i=C_{2i}x+D_{2i}\xi+{\bar D_{2i}}\xi_i, 
\end{equation}
where $\xi_i(t)\in\mathbf{R}^{m_i}$ represents the measurement disturbance
at the local sensing node $i$, and processes them locally using an
information communicated by its neighbours $j$,
$j\in\mathbf{V}_i$. Depending on the nature of the disturbances $\xi$,
$\xi_i$, the processing can be done using Kalman~\cite{Olfati-Saber-2007}
or $H_\infty$~\cite{SWH-2010,U6,LaU1} filters, both using innovations in
the measurements and the neighbours' information for feedback. To be
concrete, from now on we build the presentation around the
distributed $H_\infty$ consensus filter introduced in~\cite{U6,LaU1},
although the approach to bias attack detection proposed in this paper is
general enough to allow extensions to other types of filters in an obvious
manner.  

According to \cite{U6}, suppose the disturbances $\xi$, $\xi_i$ belong to
$L_2[0,\infty)$; this assumption suffices to guarantee that equation
(\ref{eq:plant}) has an $L_2$-integrable solution on any finite time
interval $[0,T]$, even when the matrix $A$ is unstable.     
Then using the Luenberger type observer, each filter produces an estimate $\hat x_i$ of the state $x$
\begin{eqnarray}
    \dot{\hat x}_i=A\hat x_i + L_i(y_i(t)-C_{2i}\hat x_i)+
    K_i\sum_{j\in \mathbf{V}_i}(\hat x_j-\hat x_i), 
  \label{UP7.C.d.unbiased} \\
 \hat x_i(0)=0, \nonumber
\end{eqnarray}
where the matrices $L_i$, $K_i$ are the 
parameters of the filter. The observer structure indicates that each node
takes advantage of being interconnected with other nodes in that each filter uses its neighbours estimates $\hat x_j$, $j\in \mathbf{V}_i$.   
The problem in~\cite{U6} was to determine estimator gains $L_i$ and $K_i$
in \eqref{UP7.C.d.unbiased} to ensure the filter internal stability and
acceptable $H_\infty$ attenuation of the effect which disturbances have on the
consensus performance of the filter.

\subsection{The bias attack model} 

The particular problem of interest in this paper is to consider the
situation where one or several nodes of the network of observers described
in the previous sections are subject to bias attack. While a commonly
considered situation is when the attacker interferes with the measurements
and/or communications between the nodes, here in contrast, we consider the
situation where the attacker mounts an attack on the observer
dynamics. That is, we consider the situation where in lieu of (\ref{UP7.C.d.unbiased}), some of the nodes generate their estimates according to 
 \begin{eqnarray} 
    \dot{\hat x}_i=A\hat x_i + L_i(y_i(t)-C_{2i}\hat x_i)+K_i\sum_{j\in
      \mathbf{V}_i}(\hat x_j-\hat x_i)+f_i, 
  \label{UP7.C.d} ~ \\
 \hat x_i(0)=0, \nonumber
\end{eqnarray}
where $f_i$ is the attack input. From now on, our focus is exclusively on
the network of observers (\ref{UP7.C.d}).

\begin{figure}[t]
\psfrag{R}{$f_i$}
\psfrag{nu}{$-\nu_i$}
\psfrag{Y}{$\eta_i$}
\psfrag{H}{$G_i(s)$}
\psfrag{G}{$\frac{1}{s}$}
\psfrag{+}{$+$}
\psfrag{-}{$-$}
  \centering
  \includegraphics[width=0.45\textwidth]{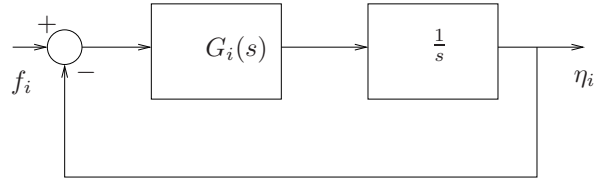}
  \caption{An auxiliary `input tracking' model.}
  \label{tracker}
\end{figure}

To present the class of admissible attack signals under consideration in
this paper, consider an auxiliary `input tracking' model shown in
Fig.~\ref{tracker}, with a stable square  
$n\times n$ transfer function $G_i(s)$, with invertible $G_i(0)$.  
Since $G_i(s)$ is square, then input $f_i$ and output $\eta_i$ of the
system in Fig.~\ref{tracker} are of
dimension $n$.
   
\begin{assumption}
  \label{A1}
Given a stable square $n\times n$ transfer function $G_i(s)$, the class of
admissible bias inputs under consideration consists of all signals
$f_i(t)$, $t\ge 0$, such that 
\begin{equation}\label{f-eta} 
 \int_0^\infty\|f_i-\eta_i\|^2dt<\infty.  
\end{equation}
\end{assumption}

Consider the tracking error of the system shown in Fig.~\ref{tracker} 
$\nu_i=\eta_i-f_i$. Under Assumption~\ref{A1}, $\nu_i$ is a finite energy
signal. Denoting the Laplace transforms of $f_i$ and $\nu_i$ 
as $f_i(s)$ and $\nu_i(s)$ respectively, and noting that
\[
\nu_i(s)=-(I_n+\frac{1}{s}G_i(s))^{-1}f_i(s),
\]
condition (\ref{f-eta}) is equivalent to 
\begin{equation}
\int_{-j\infty}^{+j\infty}
\|(I+\frac{1}{s}G_i(s))^{-1}f_i(s)\|^2ds <\infty.
\label{f-eta.Parseval}
\end{equation}
Note also that the invertibility of $G_i(0)$ guarantees that
$\lim_{t\to\infty}\|f_i(t)-\eta_i(t)\|= 0$ for inputs $f_i$ 
that have a finite limit at $\infty$.

In practice, of course the transfer function $G_i(s)$ must be selected by
the designer based on the anticipated behaviour of the attack inputs $f_i(t)$. 
It remains unknown to the attacker. For example, to capture a class of
bias injection attack inputs consisting of a steady-state component and an
exponentially decaying transient component generated by a low pass
filter~\cite{TSSJ-2015}  
it suffices to choose $G_i(s)=\frac{1}{s+2\epsilon_i}I_n$, where $I_n$ is
the $n\times n$ identity matrix, and $\epsilon_i>0$ is a constant. 
It must be noted that even with this choice of $G_i(s)$, the designer does
not need to know the asymptotic steady-state value or the shape of the
transient, as all such
signals $f_i$ satisfy condition (\ref{f-eta.Parseval}). Furthermore, such
signals have the property that $\lim_{t\to\infty}f_i(t)$ exists and
therefore we can ensure that $\|f_i(t)-\eta_i(t)\|\to 0$ as $t\to\infty$. 
More generally, signals representing a combination of constants and
$L_2$-integrable inputs satisfy  (\ref{f-eta.Parseval}). In addition to
bias attack policies $f_i$  described above, $L_2$-integrable inputs
$f_i$ are included which represent attack inputs with limited energy
resource~\cite{TSSJ-2015}.    

It can be readily shown that the state-space model for the system in
Fig~\ref{tracker} can be written as 
\begin{eqnarray}
&&\dot\omega_i = \Omega_i\omega_i+\Gamma_i \nu_i, \label{Om.sys.general} \\
&&\eta_i= [I~0]\omega_i, \qquad \omega_i(0)=0, \nonumber
\end{eqnarray}
where $\nu_i=\eta_i-f_i$ is an $L_2$-integrable input, according to
Assumption~\ref{A1}. In particular, in the special case
$G_i(s)=\frac{1}{s+2\epsilon_i}I_n$, we have $\omega_i\in \mathbf{R}^{2n}$,
and 
\begin{equation}\label{omega}
\Omega_i=\left[\begin{array}{cc} 0 & I \\ 0 & -2\epsilon_i
    I\end{array}\right], \quad \Gamma_i=\left[\begin{array}{c} 0  \\
          -I\end{array}\right].
\end{equation}

\subsection{The proposed attack detector}

The objective of the paper is to design a (distributed) attack detection
system which is capable of tracking attack inputs satisfying
Assumption~\ref{A1}. To this end, we consider 
the following outputs which summarize the information about the network
available at node $i$, and can be used by the attack detector 
\begin{eqnarray}
  \zeta_i&=&y_i-C_{2i}\hat x_i \nonumber \\
         &=&C_{2i}(x-\hat x_i) + D_{2i}\xi+{\bar D_{2i}}\xi_i, 
                                              \label{out.y} \\
  \bar\zeta_i&=&\sum_{j\in \mathbf{V}_i} (\hat x_j-\hat x_i).
                                              \label{out.c}
\end{eqnarray}
The idea behind introducing these outputs is as follows. If node $i$ is under attack, then its predicted sensor measurement $C_{2i}\hat x_i $ is expected to be biased, compared to the actual measurement $y_i$. This must lead to a significant difference between these two signals, i.e., we must expect a large energy in $\zeta_i$. Likewise, the observer under attack is expected to cause the system to deviate from the state of consensus, causing the state of the observer $i$, $\hat x_i$ to deviate from the average estimate produced at the neighbouring nodes. Thus, the disagreement variable $\bar\zeta_i$ at node $i$ is expected to differ from similar variables produced by the rest of the network. This motivates using these outputs for detecting the attack.

Let $e_i=x-\hat x_i$ be the local estimation error 
at node $i$. Using (\ref{eq:plant}) and (\ref{UP7.C.d}), it is 
straightforward to verify that the local filter errors satisfy the
following equation: 
 \begin{eqnarray}
    \dot{e}_i&=&(A - L_iC_{2i})e_i+K_i\sum_{j\in
      \mathbf{V}_i}(e_j-e_i) \nonumber \\ &+&(B_2-L_iD_{2i})\xi-L_i{\bar
      D_{2i}}\xi_i-f_i, \quad  e_i(0)=x_0.  \label{e} 
\end{eqnarray}
The outputs (\ref{out.y}), (\ref{out.c}) can be rewritten in terms of the
estimation errors as 
\begin{eqnarray}
  \zeta_i&=&C_{2i}e_i + D_{2i}\xi+{\bar D_{2i}}\xi_i, 
                                              \label{out.y.1} \\
  \bar\zeta_i&=&-\sum_{j\in \mathbf{V}_i} (e_j-e_i).
                                              \label{out.c.1}
\end{eqnarray}
Hence, we can consider the collection of systems (\ref{e}) as a large-scale
plant governed by the vector of attack inputs $f=[f_1',\ldots,f_N']'$, and
equipped with the outputs (\ref{out.y.1}), (\ref{out.c.1}). It is worth
stressing that these outputs can be readily generated at the observer $i$,
computing them only requires the local measurements $y_i$, the local
estimate $\hat x_i$ computed by the observer at node $i$ and the neighbours
estimates $\hat x_j$, $j\in\mathbf{V}_i$, available to that 
observer. Therefore the outputs (\ref{out.y.1}), 
(\ref{out.c.1}) are available for tracking the attack inputs. To achieve
this, consider the system combining the estimation error dynamics (\ref{e})
and the auxiliary input tracking model 
(\ref{Om.sys.general}):
\begin{eqnarray}
    \dot{e}_i&=&(A - L_iC_{2i})e_i+K_i\sum_{j\in
      \mathbf{V}_i}(e_j-e_i) -[I~ 0]\omega_i \nonumber \\ 
&+&(B_2-L_iD_{2i})\xi-L_i{\bar
      D_{2i}}\xi_i+\nu_i, \quad  e_i(0)=x_0,  \nonumber \\
      \dot \omega_i&=&\Omega_i \omega_i+\Gamma_i \nu_i \quad
      \omega_i(0)=0. \label{e.ext.nu.om} 
\end{eqnarray}

The system (\ref{e.ext.nu.om}) equipped with the outputs  (\ref{out.y.1}),
(\ref{out.c.1}) is an uncertain system governed by $L_2$-integrable inputs
$\xi$, $\xi_i$ and $\nu_i$. Each such system is interconnected with its
neighbours via inputs $e_j$, and the collection of all such systems
represents a large-scale system. We propose the following distributed
$H_\infty$ observer for this large-scale system which utilizes the outputs
(\ref{out.y.1}), (\ref{out.c.1}) to obtain estimates of $e_i$ and
$\omega_i$ while attenuating the disturbances $\xi$, $\xi_i$ and $\nu_i$,
$i=1,\ldots,N$: 
\begin{eqnarray}
    \dot{\hat{e}}_i&=&(A - L_iC_{2i})\hat{e}_i+K_i\sum_{j\in
      \mathbf{V}_i}(\hat{e}_j-\hat{e}_i) - [I~ 0]\hat\omega_i \nonumber \\ 
    &+&F_i(\zeta_i-C_{2i}\hat{e}_i)+H_i\left(\bar\zeta_i+\sum_{j\in
      \mathbf{V}_i}(\hat{e}_j-\hat{e}_i)\right), \nonumber \\
  \dot{\hat\omega}_i &=& 
\Omega_i \hat\omega_i +
    F^\eta_i(\zeta_i-C_{2i}\hat{e}_i)+H^\eta_i\left(\bar\zeta_i+\sum_{j\in 
      \mathbf{V}_i}(\hat{e}_j-\hat{e}_i)\right), \nonumber \\
&& \hat{e}_i(0)=0, \quad \hat\omega_i(0)=0.
\label{ext.obs.nu.1.om}
\end{eqnarray}
The coefficients $F_i$, $H_i$, $F^\eta_i$, $H^\eta_i$ are to be
found in such a way that $\hat\eta_i: \hat{\eta}_i= [I~0]\hat{\omega}_i$ tracks the output $\eta_i$ of the
auxiliary system (\ref{Om.sys.general}). Then, since according to the
definition of the auxiliary signal $\eta_i$, this signal represents $f_i$
asymptotically, we propose using $\hat\eta_i$ as a residual variable indicating
whether the attack is taking place.  

To formalize the above idea, introduce the error vectors $z_i=e_i-\hat{e}_i$,
$\delta_i=\omega_i-\hat\omega_i$. Using the extended system model
(\ref{e.ext.nu.om}) 
and the corresponding observer (\ref{ext.obs.nu.1.om}), the evolution of these
error vectors is governed by the following equations 
 \begin{eqnarray}
    \dot z_i&=&(A - L_iC_{2i})z_i+K_i\sum_{j\in
      \mathbf{V}_i}(z_j-z_i) - [I~ 0]\delta_i \nonumber \\ 
    &-&F_iC_{2i}z_i+H_i\sum_{j\in
      \mathbf{V}_i}(z_j-z_i) \nonumber \\ &+&(B_2-L_iD_{2i})\xi-L_i{\bar
      D_{2i}}\xi_i+\nu_i \nonumber \\
&-&F_iD_{2i}\xi-F_i{\bar
      D_{2i}}\xi_i, \quad  z_i(0)=x_0,  \nonumber \\
    \dot{\delta}_i&=&\Omega_i\delta_i -
    F^\eta_iC_{2i}z_i+H^\eta_i\sum_{j\in 
      \mathbf{V}_i}(z_j-z_i) \nonumber \\
&-&F^\eta_iD_{2i}\xi-F^\eta_i{\bar
      D_{2i}}\xi_i +\Gamma_i \nu_i, \quad  \delta_i(0)=0.
\label{ext.error.0} 
\end{eqnarray}
Note that we can introduce new variables $\tilde L_i=L_i+F_i$, $\tilde
K_i=K_i+H_i$, and re-write (\ref{ext.error.0}) as   
 \begin{eqnarray}
    \dot z_i&=&(A - \tilde L_iC_{2i})z_i+\tilde K_i\sum_{j\in
      \mathbf{V}_i}(z_j-z_i) - [I~ 0]\delta_i \nonumber \\ 
&+&(B_2-\tilde
    L_iD_{2i})\xi-\tilde L_i{\bar
      D_{2i}}\xi_i+\nu_i, \quad  z_i(0)=x_0,  \nonumber \\
    \dot{\delta}_i&=&\Omega_i\delta_i -
    F^\eta_iC_{2i}z_i+H^\eta_i\sum_{j\in 
      \mathbf{V}_i}(z_j-z_i) \nonumber \\
&-&F^\eta_iD_{2i}\xi-F^\eta_i{\bar
      D_{2i}}\xi_i +\Gamma_i \nu_i
, \quad  \delta_i(0)=0.
\label{ext.error} 
\end{eqnarray}

\begin{problem}[The $H_\infty$ detector design problem]\label{Prob1} 
The distributed attack detection problem under consideration in this paper
is to determine $\tilde L_i$, $\tilde K_i$, $F^\eta_i$, $H^\eta_i$ such
that the following conditions hold: 
\begin{enumerate}[(i)]
\item
The large-scale system (\ref{ext.error}) is
internally stable. Equivalently, the disturbance and attack-free 
large-scale system
 \begin{eqnarray}
    \dot z_i&=&(A - \tilde L_iC_{2i})z_i+\tilde K_i\sum_{j\in
      \mathbf{V}_i}(z_j-z_i) - [I~ 0]\delta_i ,  \nonumber \\
    \dot{\delta}_i&=&\Omega_i\delta_i -
    F^\eta_iC_{2i}z_i+H^\eta_i\sum_{j\in 
      \mathbf{V}_i}(z_j-z_i), \label{e.2} \\
	&& z_i(0)=x_0,  \quad  \delta_i(0)=0, \nonumber
\end{eqnarray}
must be asymptotically stable. 
\item 
In the presence of disturbances and attack signals, all from the class of
$L_2$-integrable signals, the system (\ref{ext.error}) achieves a guaranteed
level of $H_\infty$ filtering performance:  
\begin{equation}\label{objective.i.1}
\sup_{x_0, \mathbf{w}\neq 0}\,
\frac{\int_0^\infty\sum_{i=1}^N(\delta_i'Q_i\delta_i+ z_i'\bar Q_i z_i) dt}
{\|x_0\|^2_P+\sum_{i=1}^N\|\mathbf{w}_i\|_2^2}
\le \gamma^2, 
\end{equation}
where $Q_i=Q_i'>0$, $\bar Q_i=\bar Q_i'\ge 0$ are given matrices,
$\|x_0\|^2_P=x_0'Px_0$, $P=P'>0$ is a fixed matrix to be determined later,
$\mathbf{w}_i\triangleq 
[\xi',\xi_i',\nu_i']'$, $\mathbf{w}\triangleq[\mathbf{w}_1',\ldots,
\mathbf{w}_N']'$,  and $\gamma>0$ 
is a constant.
\end{enumerate}
\end{problem}

It follows from (\ref{objective.i.1}) that each attack detector variable
$\hat{\omega}_i$ provides an $H_\infty$ estimate of $\omega_i$. We now show
that provided Assumption~\ref{A1} holds, the output
$\hat{\eta}_i=[I~ 0]\hat\omega_i$ of the observer (\ref{ext.obs.nu.1.om})
converges to $f_i$, and hence it can be used as a residual
indicator of attack. 

\begin{lemma}\label{assympt.tracking}
Suppose Assumption~\ref{A1} holds and the observer network
(\ref{UP7.C.d}) is such that the disturbance and attack-free 
large-scale system (\ref{e.2}) is asymptotically stable, and also
(\ref{objective.i.1}) holds with $\bar Q_i>0$. Then $\|\hat\eta_i-f_i\|\to
0$ as $t\to \infty$ for all $f_i$ that have a finite limit at $\infty$.
\end{lemma}


  
Note that (\ref{objective.i.1}) with $\bar Q_i>0$ requires the observer to
ensure disturbance attenuation with respect to both $\delta_i$ and $z_i$,
even though only the variable $\delta_i$ captures the tracking error of
interest.  
When $\bar Q_i=0$ and condition  (\ref{objective.i.1}) reduces to a 
weaker condition we can guarantee that $\hat{\eta}_i$ converges to $f_i$ in
$L_2$ sense, even when $f_i$ does not have a finite limit at $\infty$.

\begin{lemma}\label{L2.tracking}
Suppose Assumption~\ref{A1} holds and the observer network
(\ref{UP7.C.d}) is such that the disturbance and attack-free 
large-scale system (\ref{e.2}) is asymptotically stable, and also condition 
(\ref{objective.i.1}) holds with $\bar Q_i=0$,
\begin{equation}\label{objective.i.2}
\sup_{x_0, \mathbf{w}\neq 0}\,
\frac{\int_0^\infty\sum_{i=1}^N\delta_i'Q_i\delta_i dt}
{\|x_0\|^2_P+ \sum_{i=1}^N\|\mathbf{w}_i\|_2^2}
\le \gamma^2. 
\end{equation}
Then $\sum_{i=1}^N\int_0^\infty \|\hat\eta_i- f_i\|^2dt<\infty$.
\end{lemma}


It is worth noting that the system (\ref{ext.obs.nu.1.om}) is governed by
the outputs of the observer network (\ref{UP7.C.d}); therefore it can be
implemented to monitor the health of the network. We explain in the next
section how to design  $\tilde L_i$, $\tilde K_i$, $F^\eta_i$, and
$H^\eta_i$ such that the above conditions hold.

\section{Attack detector design}
\label{sec:lmi_design}

Problem~\ref{Prob1} belongs to the class of distributed stabilization by
output injection problems. References~\cite{HCN-2004,U6,LaU1} developed a
vector dissipativity approach to solve this class of problems which will be
applied here as well. For each node $i$, consider a candidate storage function $V_i(z_i,\delta_i)=[z_i'~\delta_i']X_i [z_i'~\delta_i']'$, where $X_i=X_i'>0$. 
%
The following vector dissipation inequality is
instrumental in proving input tracking properties of the distributed attack
detector (\ref{ext.obs.nu.1.om}):    
\begin{equation}
\dot V_i+ 2\alpha_iV_i+ \delta_i'Q_i\delta_i+z_i'\bar Q_iz_i \le \sum_{j\in
  \mathbf{V}_j}\pi_j V_j +\gamma^2\|\mathbf{w}_i\|^2,
\label{vec.Lyap}
\end{equation}
where $\pi_i,\pi_j$ are constants selected so that the matrix
\[
\left[
 \begin{array}{cccc}
    -2\alpha_1 & \pi_2\mathbf{a}_{12} & \ldots & \pi_N\mathbf{a}_{1N}\\
    \pi_1\mathbf{a}_{21} & -2\alpha_2 & \ldots & \pi_N \mathbf{a}_{2N}\\
    \vdots & \vdots & \ddots & \vdots \\
     \pi_1\mathbf{a}_{N1} & \pi_2 \mathbf{a}_{N2}& \ldots &  -2\alpha_N
  \end{array}
\right]
\]
is diagonally dominant (and therefore it is Hurwitz~\cite{Siljak-1978}); here, $\mathbf{a}_{ij}$ is the 
element of the adjacency matrix of the graph $\mathbf{G}$. 
Indeed adding the inequalities (\ref{vec.Lyap}) will result in
\[
\begin{split}
\sum_{i=1}^N \dot{V}_i&+ \sum_{i=1}^N(\delta_i'Q_i\delta_i+z_i'\bar Q_iz_i) \\
\le & 
\max\{-2\alpha_1+q_1\pi_1, \ldots, -2\alpha_N+q_N\pi_N\}\sum_{i=1}^N V_i\\
&+
\gamma^2\sum_{i=1}^N (\|\xi\|^2+\|\xi_i\|^2+\|\nu_i\|^2).
\end{split}
\]
Selecting $\pi_i<\frac{2\alpha_i}{q_i}$ and letting
$\varepsilon=\min\{2\alpha_1-q_1\pi_1, \ldots, 
2\alpha_N-q_N\pi_N\}>0$, $V=\sum_{i=1}^NV_i$, we then have 
\begin{eqnarray}
\lefteqn{\dot V+ \sum_{i=1}^N(\delta_i'Q_i\delta_i+z_i'\bar Q_iz_i)\le} &&\nonumber \\
&& -\varepsilon V+
\gamma^2\sum_{i=1}^N (\|\xi\|^2+\|\xi_i\|^2+\|\nu_i\|^2).
\label{Lyap.global}
\end{eqnarray}
This implies that when $\xi=0$ and $f_i=0$, $\xi_i=0$ $\forall i$, then
\[
\dot V<-\varepsilon V,
\] 
and provided $X_i>0$, we have $z_i\to 0$, $\delta_i\to 0$
exponentially. That is, condition (i) of Problem~\ref{Prob1} is
established.    

Also, when at least one of the signals $\xi$,  $\xi_i$  or $f_i$ is not equal
to zero (the latter is equivalent to $\nu_i\not\equiv 0$), then it follows
from (\ref{Lyap.global}) that 
\[
\begin{split}
\sum_{i=1}^N &\int_0^T(\delta_i'Q_i\delta_i+z_i'\bar Q_iz_i)dt 
\le \sum_{i=1}^N \left[V_i(z_i(0),\delta_i(0)) \right.\\
&+ \left.\gamma^2
\int_0^T(\|\xi\|^2+\|\xi_i\|^2+\|\nu_i\|^2)dt\right].  
\end{split}
\]
Note that $V_i(z_i(0),\delta_i(0))=x_0'X_i^{11}x_0$, where $X_i^{11}$ is
the upper left block in the partition of $X_i$ compatible with the
dimensions of $z_i$ and $\delta_i$. Hence (\ref{objective.i.1}) also holds
with $P=\gamma^{-2}\sum_{i=1}^NX_i^{11}$.  
It follows
from this discussion that condition (\ref{vec.Lyap}) ensures satisfaction
of the conditions of Lemma~\ref{assympt.tracking}. Therefore, to ensure
that the distributed observer (\ref{ext.obs.nu.1.om}) can track the attack
input $f_i$ we need to determine coefficients $\tilde L_i$, $\tilde K_i$,
$F^\eta_i$, and $H^\eta_i$ for it so that (\ref{vec.Lyap}) is satisfied.
\bigskip

To present conditions under which (\ref{vec.Lyap}) holds,  introduce the
notation 
\begin{eqnarray}
  \label{notation}
&&  
A^\mu_i=\left[\begin{array}{cc} A & -[I~ 0]\\ 0 & \Omega_i 
  \end{array}\right], \quad 
B_1^\mu=\left[\begin{array}{c}I \\ \Gamma_i \end{array}\right], \quad 
B_2^\mu= \left[\begin{array}{cc} -B_2 & 0\\ 0 & 0 
  \end{array}\right], 
\nonumber \\
&&
D_{2i}^\mu= \left[\begin{array}{cc} D_{2i} & \bar
    D_{2i} \end{array}\right],\quad 
C_{2i}^\mu= \left[\begin{array}{ccc} C_{2i} & 0 
  \end{array}\right], \quad
H^\mu= \left[\begin{array}{ccc} I & 0 
  \end{array}\right], \quad \nonumber \\
&&
L_i^\mu= \left[\begin{array}{c}\tilde L_i \\ F_i^\eta \end{array}\right], \quad
K_i^\mu= \left[\begin{array}{c}\tilde K_i \\ H_i^\eta \end{array}\right]
.
\end{eqnarray}
Suppose $D_{2i}$ and $\bar{D}_{2i}$ satisfy the condition
\begin{equation}\label{E2i}
   E_{2i}\triangleq D_{2i}^\mu (D_{2i}^\mu)' = D_{2i}D_{2i}'+{\bar
     D_{2i}}{\bar D_{2i}}'> 0. 
\end{equation}
The above assumption on $E_{2i}$ is a standard  assumption made in nonsingular $H_\infty$ control problems \cite{bacsar2008h}.

\noindent Now let us introduce the matrix 
\begin{eqnarray}
  \label{Qmu}
  Q_i^\mu=\left[\begin{array}{cc} \bar Q_i & 0\\ 0 & Q_i
    \end{array} \right], 
\end{eqnarray}
where $Q_i=Q_i'>0$. Also, $\bar Q_i=\bar
  Q_i'$ is selected to be positive definite when the aim is to design
  an attack observer to achieve asymptotic tracking of attack inputs. If
  $L_2$ tracking is acceptable, one can let $\bar Q_i=0$. 
Given $\alpha_i>0$, define $\pi_i=
\frac{2\alpha_i}{q_i+1}$, where $q_i$ is the out-degree of the graph node
$i$. Clearly $\pi_i= \frac{2\alpha_i}{q_i+1}< \frac{2\alpha_i}{q_i}$.

\vspace{.2cm}
\begin{theorem}
\label{theorem}
Suppose Assumption~\ref{A1} holds and the digraph $\mathbf G$, the matrices
$Q_i=Q_i'>0$, $\bar Q_i=\bar Q_i'>0$, 
$i=1,\cdots,6$ and the constants $\alpha_i>0$, $i=1,\cdots,N$
are such that the coupled linear matrix inequalities in \eqref{LMI} (on the
next page) with
respect to the variables $X_i = X_i'>0$ and $M_i$, $i=1,\cdots,N$ are
feasible. Then choosing   
\begin{equation}\label{L}
\begin{split}
K_i^\mu &= -X_i^{-1} M_i, \\
L_i^\mu &=(\gamma^2X_i^{-1}(C_{2i}^\mu)'-B_2^\mu(D_{2i}^\mu)') E_{2i}^{-1} \\
\end{split}
\end{equation}
ensures that the condition \eqref{vec.Lyap} holds. 

\begin{figure*}[!t]
\centering
\begin{eqnarray}
\left[\begin{array}{cccccc}
S_i & X_iB_1^\mu &
X_iB_2^\mu\Big(I-(D_{2i}^\mu)'E_{2i}^{-1}D_{2i}^\mu\Big) & -M_iH^\mu & \ldots &
-M_iH^\mu \\
(B_1^\mu)'X_i & -\gamma^2 I & 0 & 0  & \ldots & 0 \\
(I-(D_{2i}^\mu)'E_{2i}^{-1}D_{2i}^\mu)(B_2^\mu)'X_i & 0 & -\gamma^2I & 0  &
\ldots & 0 \\
 -(H^\mu)'M_i' & 0 & 0 & -\frac{2\alpha_{j_1}}{q_{j_1}+1}X_{j_1} & \ldots &
 0 \\
\vdots & \vdots & \vdots & \ddots & \vdots & \vdots  \\
 -(H^\mu)'M_i' & 0 & 0 & 0 & \ldots &
 -\frac{2\alpha_{j_{p_i}}}{q_{j_{p_i}}+1}X_{j_{p_i}}
\end{array}\right]<0, \label{LMI} 
\end{eqnarray}
\begin{eqnarray}
S_i&=&
X_i\left(A_i^\mu+\alpha_i I +B_2^\mu(D_{2i}^\mu)' E_{2i}^{-1} C_{2i}^\mu\right) + \left(A_i^\mu+\alpha_i I +B_2^\mu(D_{2i}^\mu)' E_{2i}^{-1}
  C_{2i}^\mu\right)'X_i \label{S_i} \nonumber \\
&&
+p_iM_iH^\mu+p_i(H^\mu)'M_i'+Q_i^\mu - \gamma^2 (C_{2i}^\mu)'E_{2i}^{-1}C_{2i}^\mu. \nonumber
\end{eqnarray} 
\hrulefill		
\end{figure*}
\end{theorem}

Combined with
Lemma~\ref{assympt.tracking} or Lemma~\ref{L2.tracking}, this theorem provides a
complete result on the design of biasing attack detectors for the
distributed observer~(\ref{UP7.C.d}).

\section{Conclusion} \label{sec:conclusion}
The paper is concerned with the problem of distributed attack detection is sensor networks. We consider a group of consensus-based distributed estimators and assume that the estimator dynamics are under attack. Then we propose a distributed $H_\infty$ attack detector which allows for a broad range of uncertainty in the sensors and the plant model, as well as a quite broad range of bias attack inputs, and show that the proposed attack detector can track individual attack inputs at different sensors. A possible future direction is to construct a compensator to cancel the detected attack in the system.

\section*{Acknowledgement}
The authors thank G. Seyboth for providing his
paper~\cite{SA-2015}.

\newcommand{\noopsort}[1]{} \newcommand{\printfirst}[2]{#1}
  \newcommand{\singleletter}[1]{#1} \newcommand{\switchargs}[2]{#2#1}

\end{document}